# Reliability and Safety Modeling of a Digital Feed-Water Control System

Shawkat S. Khairullah

*University of Mosul, College of Engineering, Department of Computer Engineering*

Shawkat.sabah@uomosul.edu.iq

Ahmed A. Mostfa

*University of Al Hamdaniya, College of Education, Department of Computer Science*

mostfa@uohamdaniya.edu.iq

**Abstract:** *Much digital instrumentation and control systems embedded in the critical medical healthcare equipment, aerospace devices, and nuclear industry have obvious consequence of different failure modes. These failures can affect the behavior of the overall safety-critical digital system and its ability to deliver its dependability attributes if any defected area that could be a hardware component or software code embedded inside the digital system is not detected and repaired appropriately. The safety and reliability analysis of safety-critical systems can be accomplished with Markov modeling techniques which could express the dynamic and regenerative behavior of the digital control system. Certain states in the system represent system failure, while others represent fault free behavior or correct operation in the presence of faults. This paper presents the development of a safety and reliability modeling of a digital feedwater control system using Markov-based chain models. All the Markov states and the transitions between these states were assumed and calculated from the control logic for the digital control system. Finally, based on the simulation results of modeling the digital feedwater control system, the system does meet its reliability requirement with the probability of being in fully operational states is 0.99 over a 6 months' time.*

---

## I. Introduction

Digital control systems are a critical facet to the reliability and safety of Nuclear Power Plants (NPPs) operation. The control systems are continuously reading plant information and sending commands to actuating units [1]. The dependability analysis of these systems is used to increase the performance, resilience, and output capacity of the safety critical system, in particular redundant structures have been used to design robust engineering systems [1][2]. Computer-based dependable embedded systems have been designed to recognize and tolerate their own faults; i.e. fault-tolerant computer systems. For instance, Reconfiguration, the process of removing faulty components and either replacing them with spares or degrading to an alternate configuration. Although the capability of tolerating certain faults, these systems are still susceptible to different failure modes. Thus, the reliability of these systems must be evaluated to ensure that design requirements are met. Traditionally, the reliability analysis of a complex digital system consisting of many components has been accomplished using combinatorial mathematics. The standard "fault tree" method of reliability analysis is based on such mathematics. Fault tree tries to reduce the number of behavior to be considered in order to get the safe design. It analyzes the system in top-down approach trying to find the possible path of undesirable state [3]. Unfortunately, the fault-tree approach is incapable of analyzing systems where reconfiguration is possible. In reconfigurable systems the critical factor often becomes the effectiveness of the dynamic reconfiguration process. Dynamic fault free (DFT) however, is used to utilize the Architectural Analysis and Design Language (AADL) to model Digital Feed-Water Control System (DFWCS) [4]. It is necessary to model such systems using the more powerful Markov modeling technique [5]. However, one of the disadvantages of Markov model is the number of attributes (states) that represent system characteristics such as the number of working processors. The more attributes included in the model design the more complex the system will be.

Modeling a system with Markov model require representing each state of the system with their attributes, which considered system characteristics. Where, the smallest set of attributes included in the model the less complex the system will be and can describe the fault behavior. The engineering for system focus design and development efforts on the functional behavior required under normal operating conditions and given environmental assumptions such that consideration of failure scenarios could postponed until the design is completed. Thus, Safety analysis should be conducted to identify possible ways to improve the reliability and safety in the DFWCS system [6]. The next step in the modeling process is to characterize the transition time from one state to another. The system architecture used degradable triplex system. Which means transition from one state to another represents a failure with a non-constant rate. The reliability of the system is often faced many of such trade-off cases where real time critical systems should be conservative direction and represent system failure properly.

In this paper, the conservative approach will be used exclusively. All transitions in the reliability model are deducible from its architecture. This measurability is the main consideration in developing a model for a digital system. The transitions of a fault-tolerant system are divided into two categories: slow failure transitions and fast recovery transitions. If the states of the model are defined properly, then the slow failure transitions can be obtained from field data. The primary problem is to model the system so that the determination of these transitions is as easy as possible. If the model has many detailed states, the number of transitions that must be measured can be exorbitant. The paper will enlighten the expectations results of designing reliability and safety model of a digital feedwater control system. We used the computer program SURE (Semi-Markov Unreliability Range Evaluator) [7][8] to solve reliability model numerically with Markov Model to design fault tolerant Digital Feed-Water Control System (DFWCS) used to control the input water level in its associated steam generator.

## II. Overview of the Digital Feedwater Control System

Digital Feed-Water Control System (DFWCS) is used to control the input water level in its associated steam generator [9] from approximately 1% reactor output power up to 100% reactor output power. DFWCS is basically divided into four divisions: sensors module, fault-tolerant group1 module, fault-tolerant group2 module, and actuators module. Fault tolerance techniques based on redundancy management are typically used to detect the fault occurrence in digital systems. The redundancy can either be in time, or in hardware, which is called passive redundancy and used to identify the permanent faults occurring in combinational logic circuits [10].

The DFWCS contains a redundancy management scheme to organize the behavior of the system and avoid going to unsafe failure mode. For example, a duplication Main and Backup digital processing unit that works as a pair-of-comparing module, which can read various inputs through digital sensors interfaced to processors via wired connection and write control outputs to the actuating unit, is used as a fault-tolerant group1. Each one of these two processing units use at least two types of independent error detection techniques such as watch dog timers and self-testing unit which both are assumed to be non-perfect. The monitored inputs (sensors readings of the Plant) include reactor power level, steam and water flow, water temperature, and water level. The digital processor outputs number of sets of the operating positions of two valves: Main Flow Valve (MFV) and Bypass Flow Valve (BFV), and the Feedwater Pump (FWP) as shown in the overview figure Fig.1 and in the architectural design figure Fig. 2. The Main and Backup digital processors operate as a redundant pair, each of which reads the inputs and deliver the outputs. In case the Main processor fails, the Backup processor can take over the operation of the system. The outputs of both the Main and Backup processors are passed to a set of four Proportion, Integral, and Derivative (PID) Controllers. Three of these controllers assigned to the Main Flow Valve (MFV), Bypass Flow Valve (BFV), and Feed-Water Pump (FWP) respectively, while the fourth controller acting as a spare PID controller for either the MFV or BFV PID controllers, as shown in Fig. 2. If both main and backup processors fail, the PID controllers can operate in a manual mode. Loss of any PID controller with both processors failed is system loss. Loss of PID that is detectable leading to fail-safe shutdown and the loss of PID undetectable leading to fail-unsafe must be considered in the design and analysis.

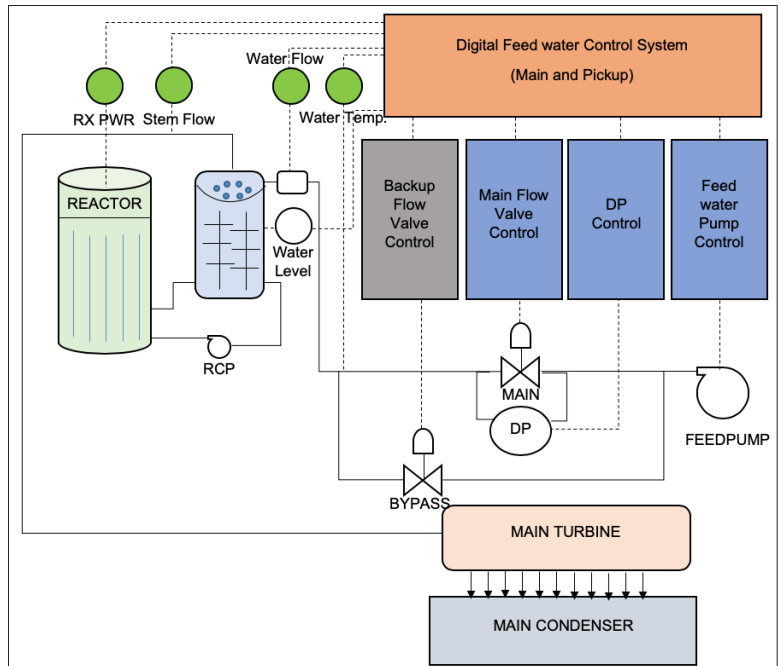

*Fig.1. Overview of the Digital Feed-Water Control System*

The PID associated with each of the controlled devices performs a "reasonableness" check of the desired setting from the Main processor and delivers the final output to the controlled device. In the event that the main processor fails, and it is detected, the backup processor takes over. The DFWCS operates in one of two primary modes: (1) Low Power Control Mode, or (2) High Power Control Mode. The first mode is used when the reactor power is less than approximately 15% and, in this mode, the MFV is normally closed, and the BFV is manipulated to control the water level in the associated steam generator. The second mode is used when the reactor power is between approximately 15% to 100% and in this mode the BFV is normally closed, and the MFV is manipulated to control the water level in the associated steam generator.

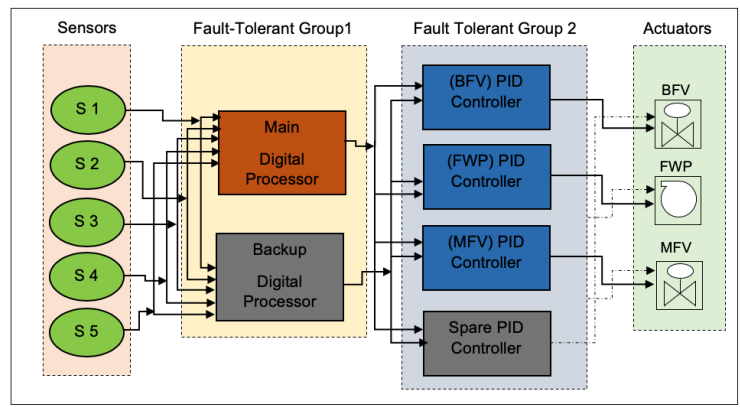

*Fig.2. System Architectural Design*

In order to increase the fault tolerance of the system, several Override Modes exist which under certain conditions allow the MFV or BFV to be used if the normally controlled valve has failed. For example, in Low Power Control Override Mode, the BFV can be bypassed and the MFV can be manipulated to control the water level. Given that tha basic parameters for the Markov model: $\lambda$ = Failure transitions. C= Error detection and reconfiguration coverage. S = State. Also, other parameters can be listed as:

$\lambda_{fwp}$ = Feed water pump failure rate
$\lambda_{FWP\_pid}$ = PID feed water pump failure rate
$\lambda_{sensor\_steam}$ = Steam sensor failure rate
$\lambda_{sensor\_WF}$ = Water flow sensor failure rate
$\lambda_{sensor\_temp}$ = Temperature sensor failure rate
$\lambda_{sensor\_level}$ = Water level Sensor failure rate
$\lambda_{sensor\_RX}$ = RX power sensor failure rate
$\lambda_{controller\_main}$ = Main controller failure = $\lambda_{controller\_backup}$ = Backup controller failure rate
$\lambda_{MFV\_pid}$ = PID_main flow valve failure rate
$\lambda_{BFV\_pid}$ = PID_backup flow valve failure rate
$\lambda_{SPARE\_pid}$ = Spare PID_valve failure rate
$\lambda_{MFV}$ = main flow valve failure rate
$\lambda_{BFV}$ =- Backup valve failure rate
$C_{controller}$ = Error detection mechanism coverage of the Main and backup processor, $C_{PID}$ = Error detection mechanism coverage of the PIDs (MFV and BFV)
μ = repair or reconfiguration rate of the main controller. (72 hours for permanent faults) (10 minutes for transient faults), $\lambda_{sensor\_Total}$ = Sum of all the sensors failure rates = $\lambda_{sensor\_steam}$ +$\lambda_{sensor\_WF}$ + $\lambda_{sensor\_temp}$ + $\lambda_{sensor\_leve}$ + $\lambda_{sensor\_RX}$

The table below shows different requirements of the altered safety integrity levels (SIL) in dependence to the probability of failure. The determination of probability of failure is extensively explained in [11].

| DFWCS Component Name | Failure Rate | Parameter (per hour) |
|---|---|---|
| Power Level Sensor | $\lambda_{PS}$ | $1 \times 10^{-6}$ |
| Steam Flow Sensor | $\lambda_{SF}$ | $1 \times 10^{-6}$ |
| Water Flow Sensor | $\lambda_{WF}$ | $1.5 \times 10^{-6}$ |
| Water Temp Sensor | $\lambda_{WT}$ | $1 \times 10^{-6}$ |
| Water Level Sensor | $\lambda_{WL}$ | $1 \times 10^{-6}$ |
| Main processor | $\lambda_{MC}$ | $3.3 \times 10^{-6}$ |
| Backup processor | $\lambda_{BU}$ | $3.3 \times 10^{-6}$ |
| Main Flow Valve PID | $\lambda_{MAIN\_FLOW\_PID}$ | $1 \times 10^{-6}$ |
| Bypass Flow Valve PID | $\lambda_{BYPASS\_PID}$ | $1 \times 10^{-6}$ |
| Spare PID | $\lambda_{SPARE\_PID}$ | $1 \times 10^{-6}$ |
| Main Flow Valve | $\lambda_{MFV}$ | $1.2 \times 10^{-6}$ |
| Bypass Flow Valve | $\lambda_{BFV}$ | $1 \times 10^{-6}$ |
| Feed-Water Pump PID | $\lambda_{FWP\_PID}$ | $1 \times 10^{-6}$ |
| Feed-Water Pump | $\lambda_{FWP}$ | $1 \times 10^{-6}$ |

### III. Design Requirements and Specifications

The design requirement of the DFWCS control system is required to be 0.99 reliable over a 6-month time. Also, the probability of being in an unsafe state should be less than $10^{-3}$. Accordingly, we classify the important states in the Markov model as operational, Fail-operational, Fail-safe, and Fail-unsafe. The markov model has been solved by using the NASA WinSURE and the NASA WinSTEM software programs. Coverage of 0.90 was assumed for both the processor coverage (Cc controller) and the PID coverage (Cp id) initially. The coverage rate was changed throughout the process of testing the model to observe the impact of the system reliability and system safety. The other specifications are listed as follows:

a. Determine the transitions between states based on the operational narrative of the digital system.
b. Assume non-perfect coverage of the error detection mechanisms. Our approach to model the system should be guided by partitioning the model into fail safe states and fail unsafe sates.
c. The proposed Markov model has used the NASA WinSure program, assume .90 coverage (initially).
d. Then, we determine the probability of being in a failsafe at the end of 400 hours and 4000 hours, and the probability of being in a fail unsafe at the end of 400, 4000 hours.

The list of the assigned symbols and coefficients used in the Markov Model is shown in table 1:

***Table 1:** Failure rates for different components for the system modeled using the Markov model shown in Fig. 3.*

| |
|---|
| $\lambda 1 = \lambda$ controller_main = 3.3x10 6. |
| $\lambda 2 = \lambda$ controller_backup =3.3x10 6 |
| $\lambda 3 = \lambda$ FWP_pid = 1x10 6 |
| $\lambda 4 = \lambda$ MFV_pid = $\lambda$ BFV_pid = $\lambda$ SPARE_pid = 1x10 6 |
| C = C controller = C pid =0.90 |
| µ permanent = Controller Repair Rate = PID Repair Rate = 1/MTTR = 1/72 hr. = 0.014 repair/hr. |
| µ transient = Controller Repair Rate = PID Repair Rate = 1/MTTR = 1/ 10 minutes. = 6 repair/hr. |

### IV. Approach used and its Implementation

To calculate the reliability and the safety of the DFWCS control system, seven states Markov model has been constructed. This model includes five operational states, one failed safe state, and one failed unsafe state as it is shown in table 2. The Markov model is intended to be illustrative of the typical parameters used in a analytical model [12], is shown in Fig. 3. The Markov model of the system is shown in Fig. 3. All the states are either in fully operational state, fail operational state, failsafe state, or the fail unsafe state. The states and the cases are illustrated in table 2:

***Table 2:** State number and status for the Markov modeled system based on fig 3.*

| State Number (1-7) | State Status |
|---|---|
| 1 | System Fully Operational (all the components are healthy) |
| 2 | Fail Operational (one processor: online or spare processor fails) |
| 3 | Fail Operational (two processors: online and spare processors fail and MFV PID used as a manual processor) |
| 4 | System Fully Operational (all the components are healthy) |
| 5 | Fail Operational (one PID: BFV or FWP PID fails) |
| 6 | Fail Safe (two PID fail with detection, two processors and MFV PID fail with detection) (System doesn't function and doesn't cause harm) failure is successfully detected by the self-diagnostic test) |
| 7 | Fail Unsafe (death) (two PID fail without detection, two processors and MFV PID fail without detection) (System doesn't function and does cause harm) failure is not detected by the self-diagnostic test) |

The model contains seven states that represents all cases that we have assumed for the system. State 1 and state 4, which both are called fully operational states; represent the case where all the components of the DFWCS control system are operating correctly. State 2, which is called fail operational state, represents the system in which either the main processor or the spare processor has failed, and that problem has been successfully detected by either the watchdog timer or the self-test process that occur inside the MFV PID. State 3, which is called fail operational state, represents the system in which both the main processor and the spare processor have failed and detected correctly. In this state, the MFV PID has used as a manual processor. State 5, which is called fail operational state, represents the system in which, either the BFV PID or the FWP PID, has failed and that problem has been detected and successfully corrected. State 6, that is called fail safe state, represents the system safely failure and controlled. A one can reached state 6 through the failure of two PIDs with successful detection due to the sequence of two faults, or through the failure of two processors and MFV PID because of the three faults sequence. State 7, which is called fail safe state, in which the system fails unsafely and cannot being controlled. A one can reached state 7 through the failure of two PIDs with successful detection due to the sequence of two faults or through the failure of two processors and MFV PID because of the tree faults sequence.

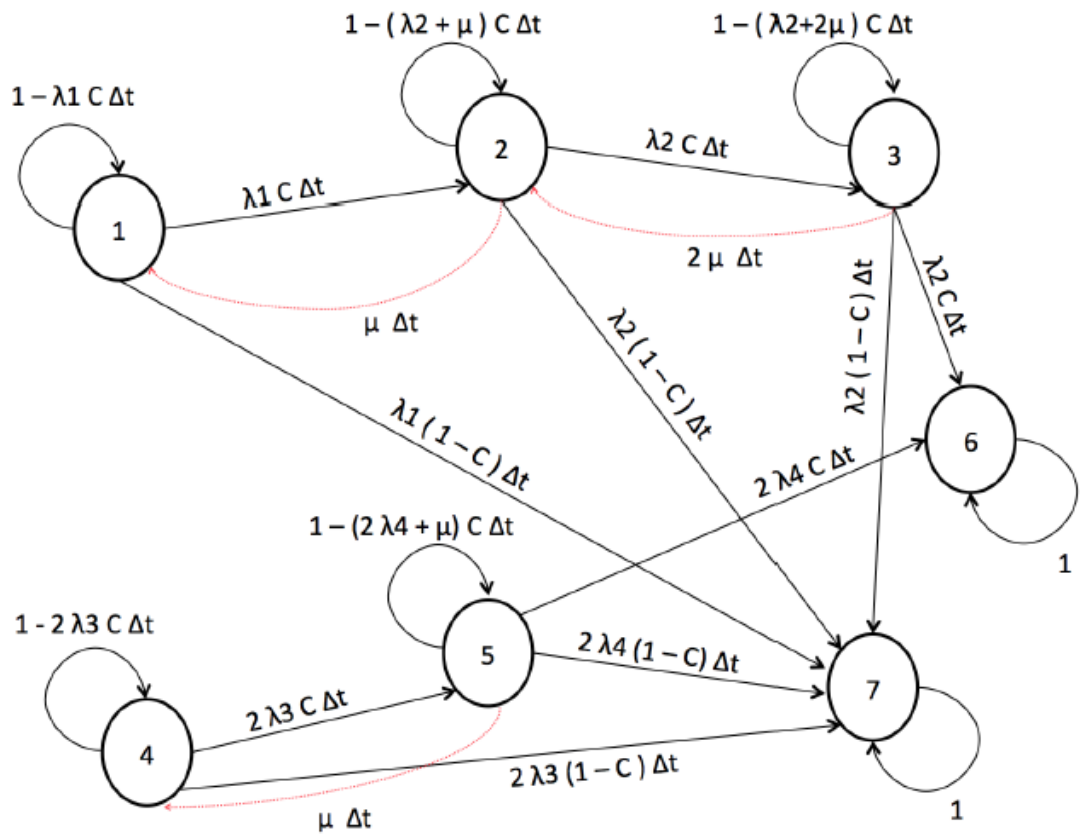

*Fig. 3. Transition State Diagram for a Digital Feed-Water Control System*

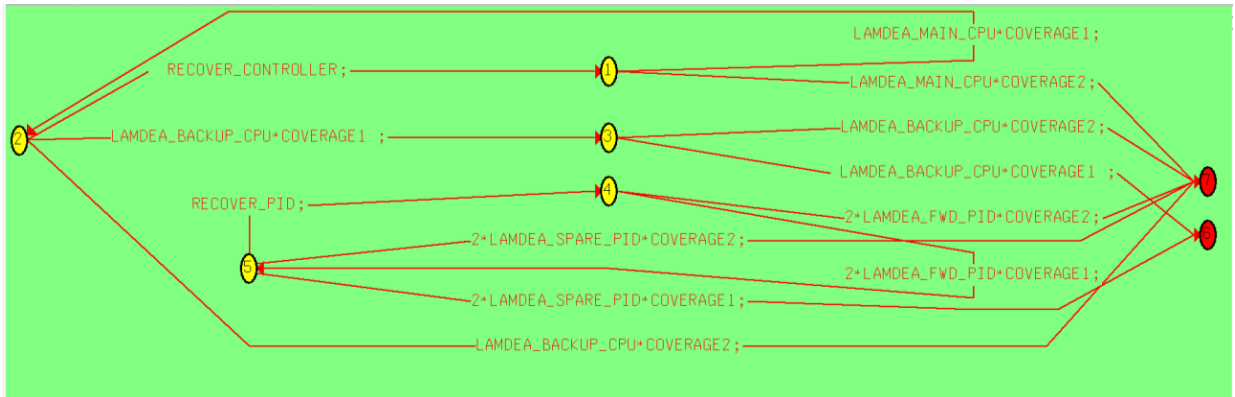

*Fig. 4. Simulation of Transition State Diagram for a Digital Feed-Water Control System using WinSURE Software program*

The circles shown in figure Fig. 4, represent the seven different states. The reference numbers (17) are represent the transition state for Markov Model in the NASA WinSURE software. The transitions arrows include the failure rates for different components, fault coverage values, and repair rates [9]. The transitions illustrated on the state diagram of Fig. 3, can be explained as follows: the transition from state 1 to state 2 can occur only if one of the two processors is failed (see Fig. 3), that failure is detected either by the watchdog timer or by the self-diagnostic of the PID, and this failure is handled correctly. Whenever this failure is not detected, the system will go into state 7, in which the system fails unsafely and may cause harm to the operators and the environment. The second transition from state 2 into state 3 is occurred only if both the online processor and the spare processor are failed and the MFV PID is used as a manual processor to keep the functionality of the system. Also, the system will go into death state (State 7), if the failure of these two processors is not detected. After the occurrence of third sequenced fault that affect on the MFV PID, the system will either go into state 6 or into sate 7 based on the successful detection of the self-diagnostic test. The transition from state 4 into state 5 can be occurred only if one PID either BFV or FWP is failed and detected correctly. After that, the system will become vulnerable to the effect of the second fault that takes the system either into state 6 or into sate 7, depending on the successful detection of this fault. The system is being repaired in three different states that are state 2, state 3, and state 5 and the repair rate in those states during Δt time is μ Δt.

The Markov model equations for the designed system can be written in matrix form, and as following:

*Psystem (t + Δt) = Tsystem * Psystem(t)*

*Where*: Tsystem: The state transition matrix.

Psystem(t): The probability of being in the corresponding state at the time t.

Psystem ( t + Δt ): The probability of being in the corresponding state at the time t + Δt.

### V. Reliability and Safety Modeling Analysis of the DFWCS

The mathematical equations of the proposed markov model of the designed system can be written from the state diagram shown in Fig. 3. [13]. For example, the probability of the system being in a state 2 at time t + Δ t depends on the probability that the system was in a state from which it could transition to state 2 and the probability of being in that state. The probability equations of the Markov model and the resulted 2D state transition matrix that includes all the failure rates and fault coverages for the system are shown below.

The probability that a system will be failed at some time (t + Δt)

P (t + Δt) = 1 - $e^{-\lambda \Delta t}$ = $\lambda \Delta t$

P1 (t + Δt) = (1 - $\lambda 1\ C\ \Delta t$) P1 (t) + μ $\Delta t$ P2 (t)

P2 (t + Δt) = (1 – ($\lambda 2$ + μ) $C\ \Delta t$) P2 (t) + 2 μ $\Delta t$ P3 (t)

P3 (t + Δt) = ($\lambda 2\ C\ \Delta t$) P2 (t) + (1 – ($\lambda 2$ + 2μ) $C\ \Delta t$) P3 (t)

P4 (t + Δt) = (1 - 2$\lambda 3\ C\ \Delta t$) P4 (t) + μ $\Delta t$ P5 (t)

P5 (t + Δt) = 2 $\lambda 3\ C\ \Delta t$ P4 (t) + (1 - (2$\lambda 4$ + μ ) $C\ \Delta t$) P5 (t)

P6 (t + Δt) = $\lambda 2\ C\ \Delta t$ P3 (t) + 2 $\lambda 4\ C\ \Delta t$ P5 (t) + P6 (t)

P7 (t + Δt) = $\lambda 1 (1 - C)\ \Delta t$ P1 (t) + $\lambda 2$ (1 - C) $\Delta t$ P2 (t) + $\lambda 2$ (1 - C) $\Delta t$ P3 (t) + 2$\lambda 3$ (1 - C) $\Delta t$ P4 (t) + 2$\lambda 4$ (1 - C) $\Delta t$ P5 (t) + P7 (t)

The equations of the discrete time Markov model for the digital feed-water control system can be written as:-

P (t + Δt) = T-system P-system (t) ➔ P ($\Delta t$) = T-system P-system (0)

The reliability of the system: R (t) = 1- P7 (t) = P6 (t) + P5 (t) + P4 (t) + P3 (t) + P2 (t) + P1 (t)

Where:-

$\frac{P1\ (t + \Delta t) - P1\ (t)}{\Delta t}$ = $\lambda 1\ C$ P1(t) + μ P2 (t)

$\frac{P2\ (t + \Delta t) - P2\ (t)}{\Delta t}$ = - ($\lambda 2$ + μ ) C P2 (t) + 2 μ P3 (t)

$\frac{P3\ (t + \Delta t) - P3\ (t)}{\Delta t}$ = $\lambda 2$ C P2 (t) - ($\lambda 2$ + 2μ ) C P3 (t)

$\frac{P4\ (t + \Delta t) - P4\ (t)}{\Delta t}$ = -2$\lambda 3$ C P4 (t) + μ P5 (t)

$\frac{P5\ (t + \Delta t) - P5\ (t)}{\Delta t}$ = 2 $\lambda 3$ C P4 (t) + -( 2$\lambda 4$ + μ) C P5 (t)

$\frac{P6\ (t + \Delta t) - P6\ (t)}{\Delta t}$ = $\lambda 2$ C P3 (t) + 2$\lambda 4$ C P5 (t)

$$\frac{P7\ (t\ +\ \Delta t) - P7\ (t)}{\Delta t} = \lambda 1\ (1 - C)P1\ (t)\ +\ \lambda 2\ (1 - C)P2\ (t)\ +\ \lambda 2\ (1 - C)P3\ (t)\ +\ 2\lambda 3\ (1 - C)P4\ (t)\ +\ 2\lambda 4\ (1 - C)P5\ (t)$$

$$\text{P-system } (t + \Delta t) = \begin{vmatrix} P1\ (t + \Delta t) \\ P2\ (t + \Delta t) \\ P3\ (t + \Delta t) \\ P4\ (t + \Delta t) \\ P5\ (t + \Delta t) \\ P6\ (t + \Delta t) \\ P7\ (t + \Delta t) \end{vmatrix}$$

The two-dimensional state transition matrix of a Markov model for the system is

$$\text{P-system } (t + \Delta t) = \begin{vmatrix} \lambda 1 C & \mu & 0 & 0 & 0 & 0 & 0 \\ 0 & -(\lambda 2 + \mu)\,C & 0 & 0 & 0 & 0 & 0 \\ 0 & \lambda 2 C & -(\lambda 2 + 2\mu)\,C & 0 & 0 & 0 & 0 \\ 0 & 0 & 0 & -2\lambda 3 C & \mu & 0 & 0 \\ 0 & 0 & 0 & 2\lambda 3 C & -(2\lambda 4 + \mu)\,C & 0 & 0 \\ 0 & 0 & \lambda 2 C & 0 & 2\lambda 4 C & 0 & 0 \\ \lambda 1\,(1 - C) & \lambda 2\,(1 - C) & \lambda 2\,(1 - C) & 2\lambda 3\,(1 - C) & 2\lambda 4\,(1 - C) & 0 & 0 \end{vmatrix}$$

$$\text{P-system } (t) = \begin{vmatrix} P0\ (t) \\ P1\ (t) \\ P2\ (t) \\ P3\ (t) \\ P4\ (t) \\ P5\ (t) \\ PFS\ (t) \\ PFU\ (t) \end{vmatrix}$$

## VI. Results and Discussions

Digital systems have multiple failure modes that depend on the detail (state space) that is embedded in modeling a system [14][15]. For the designed DFWCS control system will be completely operational as long as the system is in one of five states: state 1, state 2, state 3, state 4, and state 5 (see Fig. 3). As a result, the reliability of the system described by the seven states Markov model can be written as follows: - Reliability R (t) = P1 (t) + P2 (t) + P3 (t) + P4 (t) + P5 (t). However, the DFWCS control system will be safe if it is in one of six states: state 1, state 2, state 3, state 4, state 5, and state 6. The safety of the system described by the seven states Markov model can be written: Safety S (t) = P1 (t) + P2 (t) + P3 (t) + P4 (t) + P5 (t) + P6 (t). Reliability R (t) of a system is a function of time and is defined as the conditional probability that the system will perform its functions correctly throughout the interval [t0, t], given that the system was performing correctly at time t0. Safety S (t) of a system is defined as the probability that the system will either perform its functions correctly or will discontinue the functions in a manner that causes no harm.

When the "Plot" option was selected using the NASA WinSURE software program, the system unreliability is plotted over time. The plot shown in Fig. 5, shows that the probability of being in a death (unsafe) state after 400 hours is 0.00002 and the probability of being in a death (safe) state after 4000 hours is 0.00016. This means that the reliability is 0.9998 as it is highlighted in table 3 (fault coverage is 0.999). The requirements of the assignment required that the state remain 0.99 reliable for 6 months. Therefore, the system does meet reliability requirements.

*Table 3:* *Reliability and Safety as a function of fault coverage for the system modeled using the Markov model fig 3*

| Fault Coverage result from self-diagnostic test (C) | Reliability R (t) After 6 months | Safety S (t) | Probability of being in Fail-safe Pfs (t) | Probability of being in Fail-unsafe Pfu (t) (death) |
|---|---|---|---|---|
| 0.900 | 0.98832 | 0.98833371 | 0.00001371 | 0.01311 |
| 0.920 | 0.98949 | 0.98958664 | 0.00009664 | 0.01050 |
| 0.940 | 0.99201 | 0.99202565 | 0.00001565 | 0.00788 |
| 0.950 | 0.99330 | 0.99331616 | 0.00001616 | 0.006578 |
| 0.960 | 0.99473 | 0.994746692 | 0.000016692 | 0.005266 |
| 0.980 | 0.99734 | 0.997357772 | 0.000017772 | 0.002636 |
| 0.990 | 0.99866 | 0.998678329 | 0.000018329 | 0.001319 |
| 0.999 | 0.999849 | 0.99986784 | 0.00001884 | 0.000131 |
| 1 | 0.999981 | 0.999999898 | 0.000018898 | 0000000 |

The Probability of being in each operational state and in a death state for the digital feed-water control system is shown in Fig. 6 using WinSURE Software program. The Probability of being in each operational state and in a death state for a digital feed-water control system is shown in Fig. 7 using WinSTEM Software program.

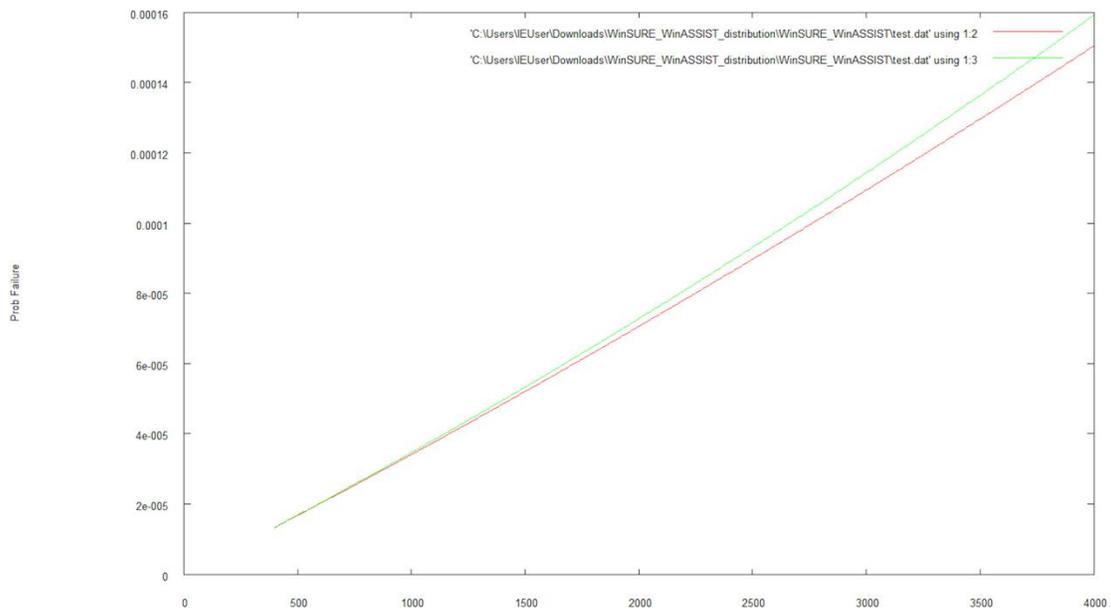

*Fig. 5.* *Probability of being in an unsafe state for a Digital Feed-Water Control System*

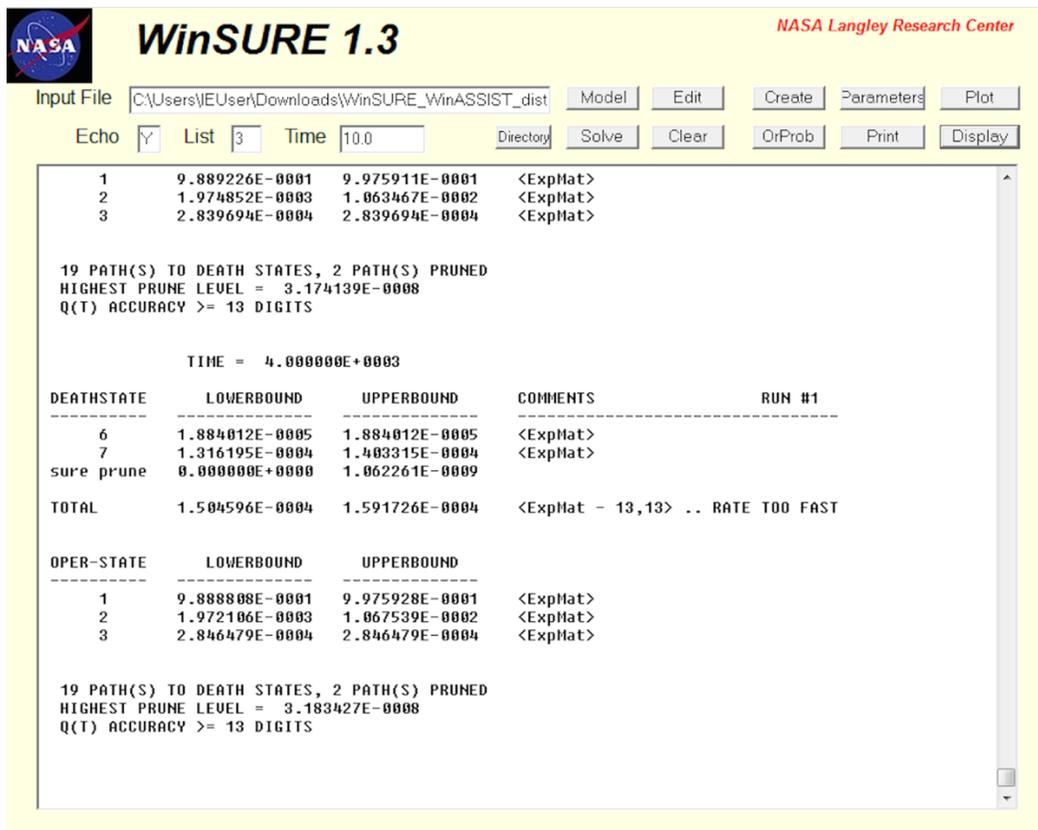

*Fig. 6.* Probability of being in operational and death states for a Digital Feed-Water Control System using WinSURE Software program

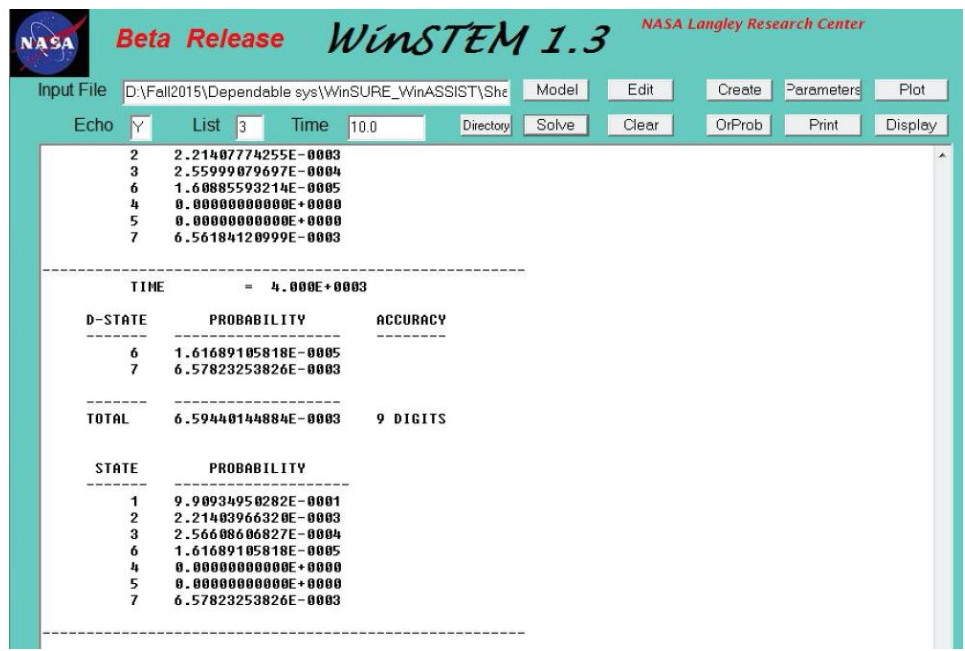

*Fig. 7.* Probability of being in operational and death states for a Digital-Feed-Water Control System using WinSTEM Software program

VII. **Conclusion**

In this paper, a conservative Markov modeling analysis for a digital feed-water control system (DFWCS) was presented. The computer program NASA WinSURE (Semi-Markov Unreliability Range Evaluator) has been used to solve reliability model numerically with Markov Model to design fault tolerant Digital Feed-Water Control System (DFWCS) used to control the input water level in its associated steam generator. The simulation results of modeling the digital control system proves that the proposed model has achieved the reliability and safety design requirements for the system with the probability of being in fully operational states is 0.99 over a 6 months' time.

**الخلاصة**: إن الكثير من الأجهزة الرقمية وأنظمة التحكم المضمنة في معدات الرعاية الصحية الطبية الحيوية ، وأجهزة الطيران ، والصناعات النووية لها نتائج واضحة لأنماط الفشل المختلفة. يمكن أن تؤثر حالات الفشل هذه على سلوك النظام الرقمي الشامل للسلامة وقدرته على تقديم سمات الموثوقية الخاصة به إذا لم يتم الكشف عن أي منطقة معيبة يمكن أن تكون مكونًا من مكونات الأجهزة أو رمز البرنامج المضمّن داخل النظام الرقمي وإصلاحه بشكل مناسب. يمكن تحقيق تحليل السلامة والموثوقية للأنظمة الهامة للسلامة باستخدام تقنيات نمذجة ماركوف التي يمكن أن تعبر عن السلوك الديناميكي والتجديدي لنظام التحكم الرقمي. تمثل حالات معينة في النظام فشل النظام ، بينما تمثل حالات أخرى سلوكًا خاليًا من الأخطاء أو عملية صحيحة في وجود أخطاء. تقدم هذه الورقة تطوير نمذجة السلامة والموثوقية لنظام التحكم الرقمي في مياه التغذية باستخدام نماذج السلسلة القائمة على ماركوف. تم افتراض جميع حالات ماركوف والتحولات بين هذه الحالات وحسابها من منطق التحكم لنظام التحكم الرقمي. أخيرًا ، استنادًا إلى نتائج المحاكاة لنمذجة نظام التحكم الرقمي في مياه التغذية ، فإن النظام يلبي متطلبات الموثوقية الخاصة به مع احتمال أن يكون في حالة التشغيل الكامل هو 0.99 على مدار 6 أشهر.